\documentclass[a4paper,fleqn,usenatbib]{mnras}

\usepackage{newtxtext,newtxmath}

\usepackage[T1]{fontenc}
\usepackage{ae,aecompl}

\usepackage{graphicx}	
\usepackage{amsmath}	
\usepackage{amssymb}	

\usepackage{graphics,subfigure,tabulary}
\usepackage[para,online,flushleft]{threeparttable}
\usepackage{caption}
\usepackage[normalem]{ulem}

\title[$H_{2}O$ and SiO masers toward V627 Cas]{Asymmetric distributions of $\rm H_2O$ and SiO masers toward V627 Cas}

\author[Yang et al.]{
Haneul Yang,$^{1,2}$
Se-Hyung Cho,$^{1,2}$\thanks{E-mail: cho@kasi.re.kr}
Youngjoo Yun,$^{2}$
Dong-Hwan Yoon,$^{2}$
\newauthor
Dong-Jin Kim,$^{3}$
Hyosun Kim,$^{2}$
Sung-Chul Yoon,$^{1}$
Richard Dodson,$^{4}$
\newauthor
Mar\'{\i}a J. Rioja,$^{4,5}$
and Hiroshi Imai$^{6}$
\\
$^{1}$Astronomy program, Department of Physics and Astronomy, Seoul National University, Gwanakgu, Seoul 08826, Korea\\
$^{2}$Korea Astronomy and Space Science Institute, Yuseong--gu, Daejeon 34055, Korea\\
$^{3}$Max-Planck-Institut f$\ddot{\textrm u}$r Radioastronomie, Auf dem H$\ddot{\textrm u}$gel 69, 53121 Bonn, Germany\\
$^{4}$International Center for Radio Astronomy Research, M468, The University of Western Australia, 35 Stirling Hwy, Crawley,\\ 
Western Australia, 6009, Australia\\
$^{5}$Observatorio Astronómico Nacional (IGN), Alfonso XII, 3 y 5, 28014 Madrid, Spain\\
$^{6}$Center of General Education, Kagoshima University, Korimoto 1-21-30, Kagoshima 890-0065, Japan
}

\date{Accepted XXX. Received YYY; in original form ZZZ}

\pubyear{2020}

\begin{document}
\label{firstpage}
\pagerange{\pageref{firstpage}--\pageref{lastpage}}
\maketitle

\begin{abstract}
We performed simultaneous observations of the $\rm H_2O$ $6_{1,6}-5_{2,3}$ (22.235080 GHz) and SiO v=1, 2, J=1$\rightarrow$0, SiO v=1, J=2$\rightarrow$1, 3$\rightarrow$2 (43.122080, 42.820587, 86.243442, and 129.363359 GHz) masers toward the suspected D-type symbiotic star, V627 Cas, using the Korean VLBI Network (KVN). Here, we present astrometrically registered maps of the $\rm H_2O$ and SiO v=1, 2, J=1$\rightarrow$0, SiO v=1, J=2$\rightarrow$1 masers for five epochs from Jan. 2016 to Jun. 2018. Distributions of the SiO maser spots do not show clear ring-like structures, and those of the $\rm H_2O$ maser are biased towards the north-northwest to west with respect to the SiO maser features according to observational epochs. These asymmetric distributions of $\rm H_2O$ and SiO masers are discussed based on two scenarios of a bipolar outflow and the presence of the hot companion, a white dwarf, in V627 Cas. We carried out ring fitting of SiO v=1, and v=2 masers and estimated the expected position of the cool red giant. The ring radii of the SiO v=1 maser are slightly larger than those of the SiO v=2 maser, as previously known. Our assumption for the physical size of the SiO maser ring of V627 Cas to be the typical size of a SiO maser ring radius ($\sim$ 4 AU) of red giants yields the distance of V627 Cas to be $\sim$ 1 kpc.
\end{abstract}




\section{Introduction}
Symbiotic stars are interacting binary systems composed of a cool red giant and a hot dwarf companion. In the symbiotic star, the interaction between the host and the companions cause complex phenomena (e.g., accretion disks around the hot companion, jets and outflows etc.). These phenomena provide clues for solving the mysteries of how spherically mass-losing red giants evolve to asymmetric bipolar or multi-polar planetary nebulae. Moreover, the symbiotic stars are considered to be potential progenitors of Type Ia supernovae \citep{cor2003, dil2012}. In order to precisely estimate a history of a mass accretion rate, it is crucial to trace the interactions between the cool star and the hot dwarf, and to investigate the morphology and dynamics of the dusty circumstellar envelope of the cool companion, especially in relation to research on the evolutionary phase from asymptotic giant branch (AGB) stars to planetary nebulae.

OH, $\rm H_2O$ and SiO masers will be useful tools for investigating such binary interactions if the cool companions are these maser sources. The SiO masers, due to the high excitation temperature and density, are suitable for investigating nearby regions of the cool companion \citep{dia1994}. On the other hand, 22 GHz $\rm H_2O$ masers are good tracers for the outflow regions above the dust forming layer \citep{reid1997}. The simultaneous observations of SiO and $\rm H_2O$ masers will allow us to trace the effect of the hot companion on the atmosphere and circumstellar envelope of the cool companion together with OH masers. 
In addition, observations of OH masers trace the outermost region of the circumstellar envelope beyond the 22 GHz $\rm H_2O$ maser region in the cool companion. Recently it has become possible to carry out high precision bona fide astrometric registration of the spatial distributions of multi-transition SiO and 22 GHz $\rm H_2O$ masers using KVN and Source Frequency Phase Referencing (SFPR) method. This is the basis of the KVN Key Science Project of evolved stars (\url{https://radio.kasi.re.kr/kvn/ksp.php}).

Long-term single dish monitoring of 22 GHz $\rm H_2O$ and 1612 MHz OH maser lines toward V627 Cas was performed by \citet{ash2017}.  The activity of the $\rm H_2O$ maser over the range of 2.8-6.0 years was monitored, providing evidence that V627 Cas is an irregular variable. Both SiO (v=1, 2, J=1-0) and $\rm H_2O$ masers were detected using the KVN single dish telescope \citep{cho2010}. However, until now, Very Long Baseline Interferometry (VLBI) observations of these masers in V627 Cas were not carried out.  

Here, we focus on VLBI observations towards V627 Cas, which emits both relatively strong $\rm H_2O$ and SiO maser lines. V627 Cas is classified as a suspected D-type (dust-rich) symbiotic star \citep{bel2000} characterized by thick dusty envelopes \citep{ber1988, kol1991}. The nature of the binary system including the companion separation and orbital period etc. is not well known. The cool component is an M2-4II spectral type star with an irregular pulsation period of about 466 days, and its average brightness is about 12 mag with sporadic one-day timescale variations, indicating the possible presence of flickering \citep{gro2006}. The activity of V627 Cas was suggested to be of a similar nature of the hot companion in the symbiotic star CH Cyg \citep{gro2006}. In this paper, we present the pilot results of the astrometrically registered maps of $\rm H_2O$ and SiO masers toward V627 Cas using the KVN. Section 2 describes the observations and data reduction. Sections 3 and 4 present the results and discussion, respectively. Finally, we summarize in Section 5.

\begin{figure*}
\centering
\includegraphics[scale=0.23]{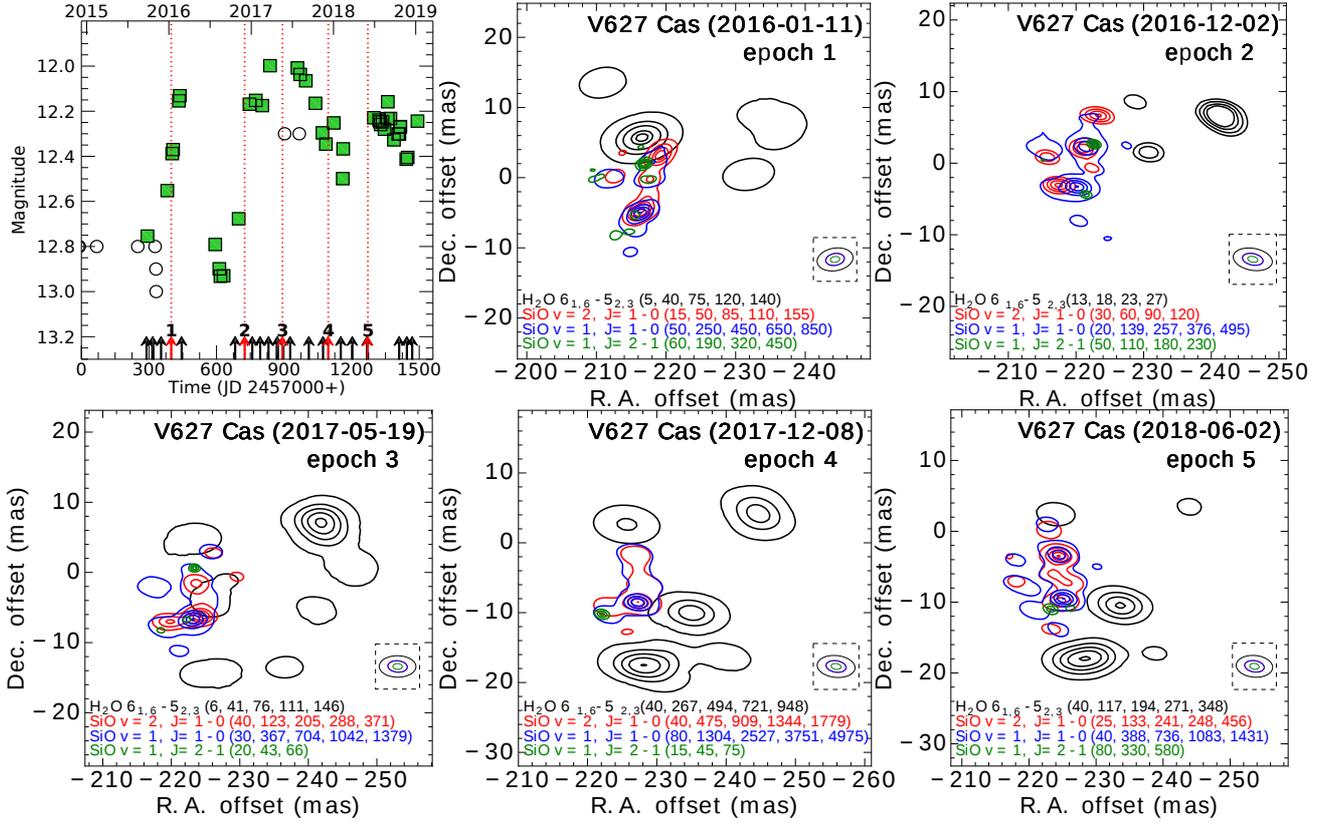}
\caption{
\textit{Upper-Left panel}: The Visual-band (open circle) and V-band (green square) light curve of V627 Cas from Dec 2014 to Jan 2019. The horizontal-axis indicates the Julian date and the vertical-axis indicates the V-band magnitude obtained from AAVSO. The arrows indicate the date of VLBI observations with KVN and the red vertical dotted lines and arrows indicate the five epochs presented in this paper. The numbers marked on the arrow indicate epoch 1 (Jan 11, 2016) , epoch 2 (Dec 2, 2016), epoch 3 (May 19, 2017), epoch 4 (Dec 8, 2017) and epoch 5 (Jun 2, 2018).
\textit{Upper-Middle to Lower-Right}: The astrometrically registered, velocity integrated intensity maps of $\rm H_2O$ (black) and SiO v=1,2, J=1$\rightarrow$0, SiO v=1, J=2$\rightarrow$1 masers (blue, red, and green, respectively) toward V627 Cas. The spatial distributions at all maser frequencies are astrometrically aligned. The peak intensity values of the masers are $3.06$/$18.90$/$3.78$/$8.66$, $0.53$/$9.28$/$2.21$/$2.36$, $1.48$/$23.14$/$4.98$/$0.57$, $3.29$/$34.24$/$15.31$/$1.47$, and $3.39$/$17.71$/$8.00$/$\rm 3.27 ~mJy~beam^{-1}~km~s^{-1}$ at each epoch, respectively. The rms noise levels on the map are $20.33$/$22.03$/$22.81$/$14.01$, $14.19$/$18.35$/$17.18$/$9.81$, $9.47$/$16.66$/$13.08$/$7.47$, $3.44$/$6.90$/$8.56$/$16.98$, $9.47$/$12.29$/$17.16$/$\rm 5.44 ~mJy~beam^{-1}~km~s^{-1}$, respectively. The contour levels are adopted by the values exhibited on the right of maser transition in the panels for multiples of each rms noise level. The synthesized beams of each maser transitions are shown in the lower right of each panel within the dotted square.
\label{fig:figure1}}
\end{figure*}

\section{Observations and data reduction}

We have performed simultaneous VLBI monitoring of $\rm H_2O$ $6_{1,6}-5_{2,3}$ (rest frequency of 22.235080 GHz) and SiO v=1, 2, J=1$\rightarrow$0, SiO v=1, J=2$\rightarrow$1, 3$\rightarrow$2 (43.122080, 42.820587, 86.243442, and 129.363359 GHz) maser lines toward V627 Cas from 2015 Sep. to 2018 Dec. (total 32 epochs) using the KVN. This array is equipped with a unique quasi-optics for simultaneous observations of K/Q/W/D bands \citep{han2008}. The advantage of this system is that it can provide images that are astrometrically registered across the different frequencies. The synthesized beam sizes are typically 6/3/1.5/1 mas in the K/Q/W/D bands, respectively.

The signal was recorded on the Mark 5B (MK5B) recorder with a data rate of 1 Gbps.
We observed two fringe finders, J2232+1143 and 3C84, and continuum delay calibrator, J2231+5922. The angular separation between V627 Cas and J2231+5922 is 3.43 degrees. Observations were performed interleaving 2 min scans on V627 Cas and continuum delay calibrator at all frequencies simultaneously. The fringe finder calibrator was observed every hour to solve for the instrumental delay and for bandpass calibration. Total observation time was about 7 hours for each epoch. The Distributed FX (DiFX) software correlator was used to correlate the recorded signal with 512 spectral channels for each base-band, yielding a velocity channel spacing of 0.42, 0.22, 0.11, and $\rm 0.07 ~km~s^{-1}$ for line observations of the K, Q, W, and D bands respectively. 

Data reduction was performed using the NRAO Astronomical Image Processing System (AIPS) package. We used the SFPR technique to obtain the astrometric registration of the $\rm H_2O$ and SiO multi transition maser spots. The basis of the SFPR technique is presented in \citet{rio2011}, and the application for the maser lines toward evolved stars is presented in \citet{yoon2018} and \citet{kim2018}. The positions of the maser spots in the maps were measured using the two-dimensional Gaussian fitting task in AIPS. 

Among the observed 32 epochs data, we present, in the visual light curve of the upper left panel of Fig \ref{fig:figure1}, the five representive epochs, which successfully provided images cubes of $\rm H_2O$, and SiO v=1, 2, J=1$\rightarrow$0, SiO v=1, J=2$\rightarrow$1 in this paper. The SiO v=1, J=3$\rightarrow$2 maser cubes could not be obtained in any epoch, due to the intensity being too weak.

\begin{figure*}
\centering
\includegraphics[scale=0.45]{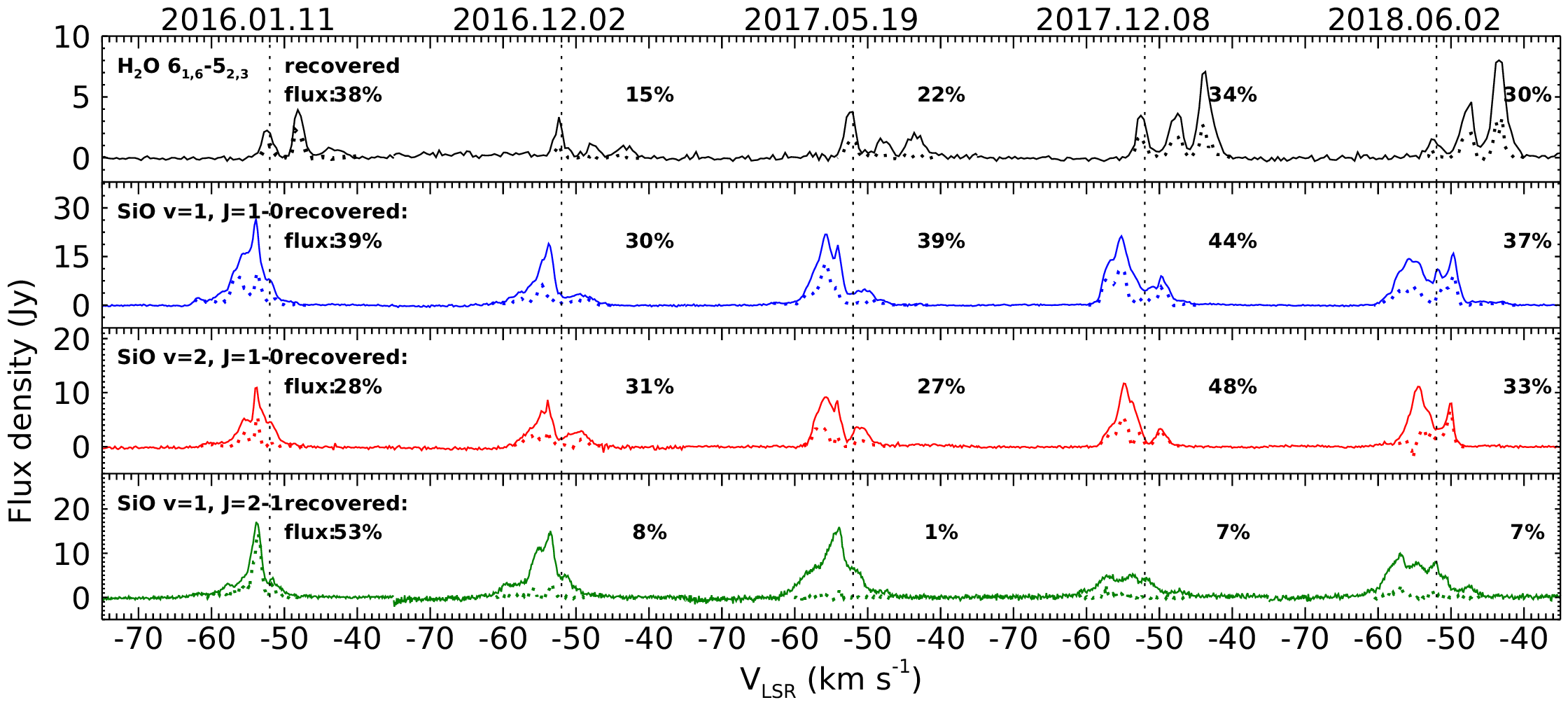}
\caption{
The total power (solid) and correlated flux (dotted) spectra of the $\rm H_2O$ and SiO masers in the 5 epochs. The total power spectra were obtained from the KVN Yonsei (epochs 1, 4, and 5) and Tamna telescopes (epochs 2 and 3). The fractions of recovered flux are marked in the top-right of the spectra. Vertical dotted lines indicate the stellar velocity of V627 Cas ($\rm V_{LSR}=-52.0~km~s^{-1}$).
\label{fig:figure2}}
\end{figure*}

\section{Results}
Fig. \ref{fig:figure1} shows the astrometrically registered, velocity-integrated, intensity maps of $\rm H_2O$ and SiO masers of these five epochs. In general, SiO masers around an isolated AGB stars exhibit ring-like shapes in the velocity-integrated maps \citep[e.g][]{yoon2018, kim2018}, which is well correlated to a uniform spatial distribution with tangential amplification. In the case of V627 Cas, the distribution of the SiO masers deviate from a ring-like shape showing gaps in the maser distributions. The SiO v=1 and v=2 masers of V627 Cas are distributed along a straight line in the epoch 1, while the distributions in the epochs 3, 4, and 5 are in arc-like shapes. The distribution of the SiO v=1, J=2$\rightarrow$1 maser also showed an arc-like feature in epochs 1, 2, and 3. 

On the other hand, the spatial distributions of the $\rm H_2O$ masers are very asymmetric and show a large time-variation from epoch 1 to epoch 5. The position at which the $\rm H_2O$ maser features are the brightest, tends to shift from north to north-west, and south-west (with respect to the SiO maser region) along the observed epochs. In all the five epochs, the $\rm H_2O$ maser features were not detected in the eastern part of the SiO maser features. In epoch 1, the $\rm H_2O$ maser features were located in the north and northwest of the SiO maser region. In epoch 2, the northern $\rm H_2O$ maser feature disappeared and only the northwestern $\rm H_2O$ maser feature was radiating. In epoch 3-5, the $\rm H_2O$ maser feature appeared in the south-west and south relative to the SiO maser region, in addition to the northwestern feature as discussed for epoch 2. The south-western $\rm H_2O$ maser features gradually develop in intensity from epoch 3 to epoch 5.

Fig. \ref{fig:figure2} shows the total power (solid) and correlated flux (dashed) spectra of the $\rm H_2O$ and SiO masers for the five epochs. The total power spectra were obtained from the KVN Yonsei (epochs 1, 4 and 5) and Tamna (epochs 2 and 3) telescopes. The fractions of recovered-flux range from 15\% to 48\% for the $\rm H_2O$ and SiO v=1, 2, J=1$\rightarrow$0 masers.  However, the fraction of the SiO v=1, J=2-1 maser ranges from 53\% to 1\%. 
The SiO maser had a velocity offset range from $\rm -10~km~s^{-1}$ to $\rm +4~km~s^{-1}$ with respect to the stellar velocity of V627 Cas, $\rm V_{LSR}=-52~km~s^{-1}$ in the local standard of rest (LSR). We also note that the SiO v=1 maser spots had the largest offset from LSR ($\rm -10~km~s^{-1}$) in epoch 5. The SiO masers peak at blueshifted velocities in all five epochs, and the peaks at the systemic and redshifted velocities grow in intensity with time, from epoch 1 to epoch 5.
On the other hand, the spectra of $\rm H_2O$ masers exhibit peak centers located at the systemic and redshifted velocities, i.e., in any of the five epochs, no $\rm H_2O$ maser peak is found with a peak center velocity smaller than $\rm -52~km~s^{-1}$. The $\rm H_2O$ maser spots had a velocity offset range from $\rm -2~km~s^{-1}$ to $\rm +12~km~s^{-1}$.

\begin{figure*}
\centering
\includegraphics[scale=0.625]{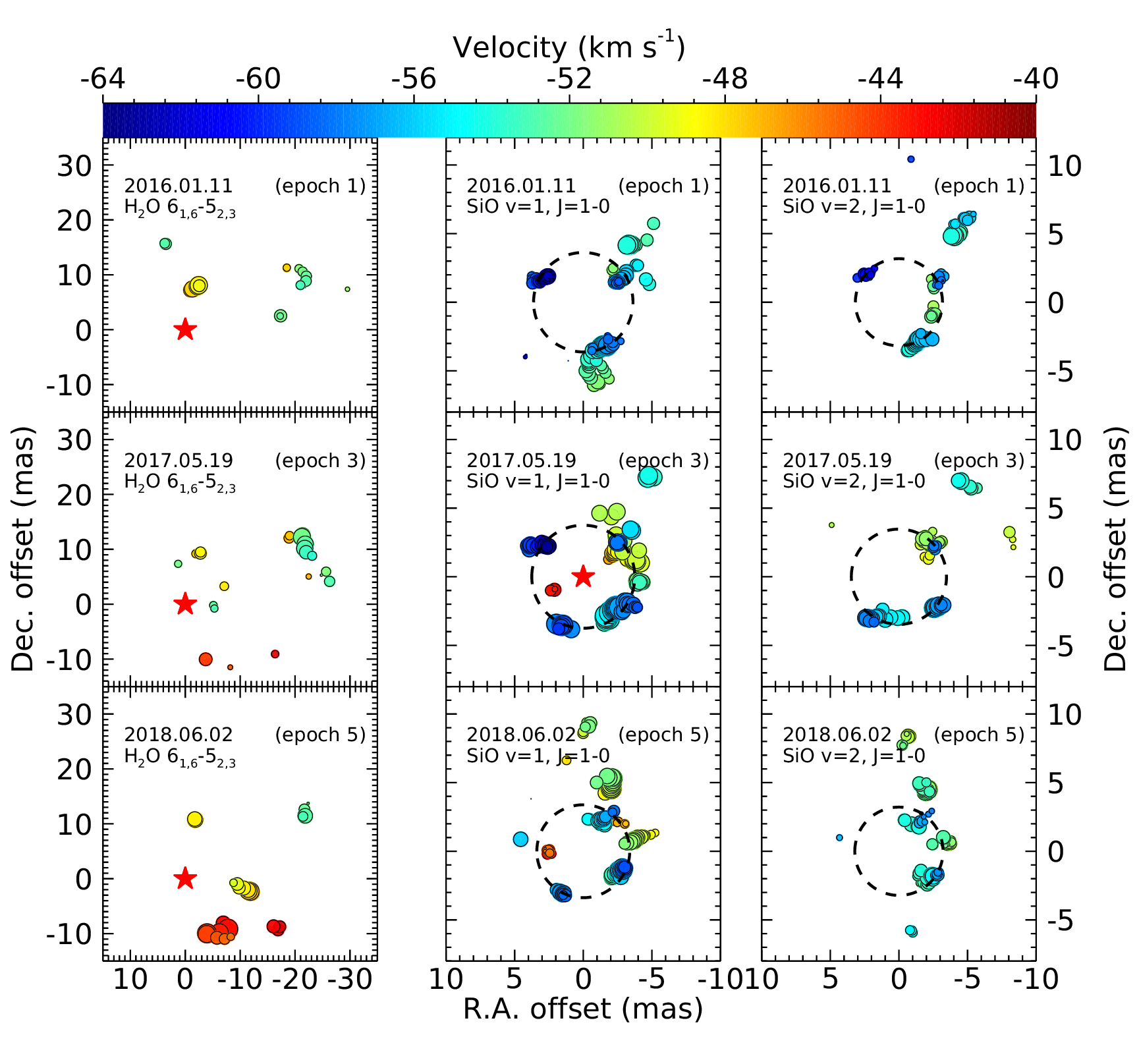}
\caption{Astrometrically registered position-velocity maps of the multiple maser lines: $\rm H_2O$ (leftmost), SiO v=1, J=1$\rightarrow$0 (middle), and v=2, J=1$\rightarrow$0 (rightmost column) masers in epochs 1, 3, and 5. The colors of the spots indicate the local-standard of rest velocity with respect to the stellar velocity according to the color bar, and the sizes of the spots indicate the logarithmic scale of intensity. The dashed circles indicate the ring fitting for the SiO v=1, and v=2 maser spot distributions for each epoch. The red star indicates the estimated position of the red giant obtained using the ring fitting results for the spot distributions of the SiO v=1 maser in epoch 3.
\label{fig:figure3}}
\end{figure*}

In Fig. \ref{fig:figure3}, we present the spatial distributions of the $\rm H_2O$ and SiO v=1, 2 maser spots with the velocity information in colors for representative epochs 1, 3, and 5 from 2016 Jan to 2018 Jun. In the case of the $\rm H_2O$ masers, no large red-shifted maser spots in south-west with respect to the red giant (red star) are found in epoch 1. They appear in the south-west from epoch 3, and show strong features in epoch 5. They coexist with the blue-shifted maser spots in epochs 3 and 5, showing an one-sided conical outflow. As shown in the velocity-integrated intensity maps in Fig. \ref{fig:figure3}, the distributions of the SiO maser spots clearly show a divergence from a ring-like shape. Only the SiO v=1 maser distributions in epoch 3 might be considered as a ring shape, if the strong north-western component is excluded (see the middle panel of Fig. \ref{fig:figure3}). These results, of the divergence from a ring-like shape including gaps in SiO maser distributions, are also shown in other previous observations. The SiO maser features in Mira variable R Cas are dominant on the eastern side during the first pulsation cycle and dominant in the western side during the remaining epochs \citep{as2011}. Long-term monitoring observations toward TX Cam show that the morphology of SiO maser emission usually resembles a ring-like structure or an ellipse with occasional deviation due to localized phenomena \citep{gon2013}. The SiO masers from IK Tau always appear distributed in an ellipse with many gaps \citep{cotton2010}.

For the majority of the large numbers of isolated single red giant stars, the SiO maser spots are arranged in a ring with its center at the stellar position, which allows for an estimation of the stellar position through the ring fitting of the maser spots. To estimate the position of red giant star in V627 Cas, we performed a ring fitting of the SiO maser spots for all the five epochs limited to the region where the maser spots clearly exhibited the ring structure in epoch 3. 
The ring fitting was carried out using IDL function, {\sc mpfitellipse} \citep{mark2009} and the results are shown in Table 1. The ring radii of the SiO v=1 maser spots are slightly larger than those of the SiO v=2 maser, as previously known. Through the ring fitting, we estimated the stellar coordinates given in Table 1. The position errors of the central star estimated by the ring-fitting of SiO masers will be composed of various factors, for example, coordinate errors of KVN telescopes, errors of atmospheric model applied in DiFX, errors of SFPR technique and ring fitting method, etc. These errors were discussed by \citet{dod2014}, \citet{yoon2018} and \citet{kim2018}.

\begin{figure*}
\centering
\includegraphics[scale=0.6]{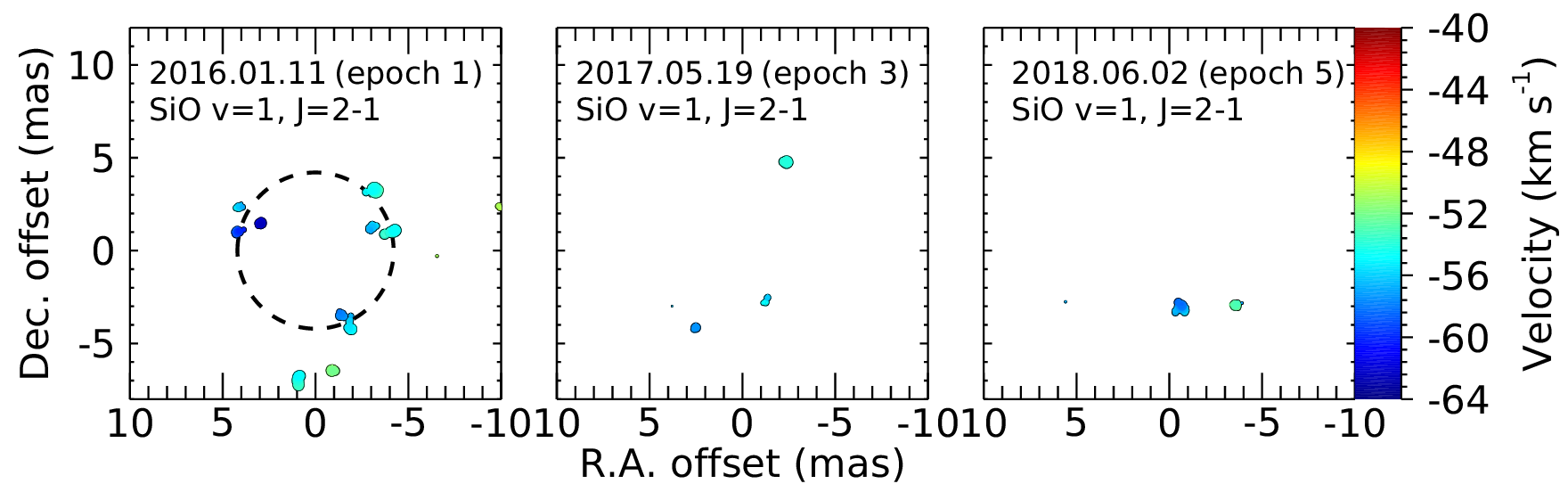}
\caption{
The position-velocity maps of the SiO v=1, J=2-1 maser in epoch 1, 3, 5 and ring fitting result in epoch 1. The color, size of spots and dashed line indicate the same features described in Fig. 3.
\label{fig:figure4}}
\end{figure*}

We were able to obtain the position-velocity maps of SiO v=1, J=2-1 maser for five epochs, but we failed to ring fit except in epoch 1, because the SiO v=1, J=2-1 maser had a limited number of spots due to the lower recovered flux ratios in the other epochs (Fig. \ref{fig:figure2}). Fig. \ref{fig:figure4} shows the position-velocity maps in epochs 1, 3, 5 and ring fitting result for the SiO v=1, J=2-1 maser in epoch 1. The ring radius in epoch 1 is about $4.20\pm0.09$ mas and is also larger than that of the SiO v=1 maser, which is consistent with the previous VLBI results for WX Psc \citep{sor2004}, and R Leo \citep{sor2007}.

\section{Discussions}

\subsection{Asymmetric distributions of $\rm H_2O$ and SiO masers in V627 Cas} \label{subsec:ringfit}

As shown in Figs. 2 and 3, the $\rm H_2O$ maser features were not detected on the eastern side of the red giant star during our observations from 2016 Jan. to 2018 Jun. In addition, the position-velocity maps of the $\rm H_2O$ maser in Fig. 3 showed an one-sided conical outflow to the the western direction. In the case of SiO masers, they did not show a clear ring-like structure and instead, significant gaps of maser features in the eastern side of the red giant. The VLBI maser features will be influenced by local conditions such as density, turbulence and temperature and the amplification path length \citep{phi2001}. Based on our VLBI results for the suspected symbiotic star V627 Cas, we can discuss two possible scenarios on the asymmetrical distributions and variations of the $\rm H_2O$ and SiO masers including the gaps.

The first scenario is a bipolar outflow around V627 Cas. We detected the one-sided conical outflow to the western direction in the $\rm H_2O$ maser maps in Fig. 3. We also found the significant gaps in SiO maser features on the eastern side compared to those of the western side. The eastern outflow of the $\rm H_2O$ maser may not be detected because of diffuse gas in this region. In the case of the Mira variable TX Cam for five epochs of SiO maser maps with the VLBA, there are clear gaps in the ring caused by a bipolar outflow \citep{gon2013}. This has been seen for other single stars, for example, IK Tau observed with MERIN \citep{bains2003}. Their channel maps of blueshifted features to the west tends to be brighter than those to the east, while redshifted features are the opposite. For the semiregular variable R Crt, the $\rm H_2O$ maser also show a very asymmetric one-sided outflow to the southern part with respect to a ring-like structure of SiO masers \citep{kim2018}.

The second scenario is the influence of the white dwarf on the spatial distributions of the $\rm H_2O$ and SiO masers around the red giant star in V627 Cas, since it is classified as a suspected D-type symbiotic star \citep{bel2000}. \citet{sea1995} carried out surveys at high sensitivities for 1612 MHz OH and 22 GHz $\rm H_2O$ masers in D-type symbiotic Miras. They found that symbiotic Miras show significant disruption of the maser structures, compared to isolated Miras. They attributed the disruption to the photodissociation of molecules by the UV photons emitted from the hot companion. The $\rm H_2O$ maser arises farther out from the cool red giant (about 10$\sim$20 stellar radii) compared to the SiO masers (about 2-4 stellar radii). Therefore, the spatial distribution of $\rm H_2O$ maser can significantly depend on the position and direction of the hot white dwarf compared to that of SiO maser. Based on this scenario, we can predict that the white dwarf was located in the east of the red giant in the V627 Cas system. The UV radiation from the hot companion will photodissociate a large number of molecules in the circumstellar envelope of the cool giant star, causing the non-detection of the $\rm H_2O$ maser to the eastern side of the red giant V627 Cas (Fig. \ref{fig:figure3}). On the other hand, the $\rm H_2O$ maser can survive on the far side of the red giant (west of the red giant center). In addition, we can speculate that the variations in $\rm H_2O$ maser spot distributions from north-southwest to west according to epochs 1, 3 and 5 are associated with the orbital motion of the white dwarf from southeast to east with respect to the red giant center. However, the interval from epoch 1 to epoch 5 is too short for comparing their variations with the orbital period of order of a few tens of years in D-type symbiotic star. Instead, these time-scale variations can be associated with the lifetime of $\rm H_2O$ maser around the red giant according to variations of local conditions.  Studies of AGB stars show that the lifetime of individual $\rm H_2O$ maser features span range from a few months to a couple years \citep{ric1999, yates1994}.

The SiO maser emitting region, which is closer to the central star than the $\rm H_2O$ maser region, is relatively opaque to the UV photons from the hot white dwarf. Therefore, the observed asymmetric distribution of SiO maser is difficult to explain in the same manner as that of the $\rm H_2O$ maser. As an alternative scenario, we can propose that the SiO masers emitting from the region facing the red giant V627 Cas is affected by a wind interaction zone. A C-facing wind interaction zone can be formed between the cool and hot stellar components when the hot companion wind has access to the polar direction \citep{ken2005}. The flow in the wind interaction zone is turbulent, with an outflow from the apex \citep{hin2013}. Assuming the components of the symbiotic systems have an equal mass-loss rate and the wind-velocity of the hot dwarf is 100 times larger than that of the cool star, \citet{hin2013} suggested that the wind interaction zone can penetrate to the SiO maser region and SiO maser radiation few examples of SiO maser spots on the east side of the red giant in V627 Cas may be related to this scenario. However, it is not yet clear which of the two scenarios will be the better explanation for asymmetric distributions of masers in V627 Cas. Further interferometric monitoring observations at higher angular resolution and sensitivity is required in order to resolve between these scenarios and to confirm the binary symbiotic system of V627 Cas.

\begin{table*}
\begin{center}
\begin{threeparttable}[h]
\caption{The radius of the SiO v=1, and v=2 maser regions and estimated coordinates of the red giant in V627 Cas.}
\label{tab:table1}
\begin{tabular*}{16.5cm}{ccccccc}
\hline \hline \noalign {\smallskip}
\textbf{Epoch} & \textbf{SiO transition} & \textbf{Ring radius} & \multicolumn{2}{c}{\textbf{Center of the ring fitting \tnote{*}}} & \multicolumn{2}{c}{\textbf{Converted coordinate (J2000)}}\\ 
& \textbf{(J=1$\rightarrow$0)} & \textbf{(mas)} & \textbf{R.A. (mas)} & \textbf{Dec. (mas)} & \textbf{R.A. (h:m:s)} & \textbf{Dec. (d:m:s)}\\
\hline\noalign {\smallskip}
1& v=1 & $3.62\pm0.21$ & $-215.04\pm0.29$  & $-1.60\pm0.34$ & 22:57:40.96231 & 58:49:12.49840\\
(Jan 11, 2016)& v=2 & $3.17\pm0.29$ & $-215.25\pm0.40$  & $-1.54\pm0.42$  & 22:57:40.96228 & 58:49:12.49846\\
\hline
2& v=1 & $3.80\pm0.22$ & $-218.87\pm0.30$  & $0.17\pm0.29$ & 22:57:40.96182 & 58:49:12.50002\\
(Dec 2, 2016)& v=2 & $3.44\pm0.24$ & $-219.22\pm0.35$ & $-0.04\pm0.40$ & 22:57:40.96177 & 58:49:12.49995\\
\hline
3& v=1 & $3.75\pm0.18$ & $-221.11\pm0.25$  & $-4.01\pm0.30$   & 22:57:40.96153 & 58:49:12.49599\\
(May 19, 2017)& v=2 & $3.46\pm0.35$ & $-221.68\pm0.60$ & $-3.77\pm0.44$ & 22:57:40.96146 & 58:49:12.49623\\
\hline
4& v=1 & $3.89\pm0.51$ & $-224.28\pm0.58$  & $-6.12\pm0.47$ & 22:57:40.96112 & 58:49:12.49388\\
(Dec 8, 2017)& v=2 & $3.35\pm0.68$ & $-225.21\pm0.60$ & $-5.85\pm0.74$ & 22:57:40.96100 & 58:49:12.49415\\
\hline
5& v=1 & $3.38\pm0.31$ & $-222.73\pm0.41$  & $-7.82\pm0.36$ & 22:57:40.96132 & 58:49:12.49217\\
(Jun 2, 2018)& v=2 & $3.22\pm0.54$ & $-222.63\pm0.65$ & $-7.73\pm0.49$ & 22:57:40.96133 & 58:49:12.49227\\
\hline \noalign {\smallskip}
\end{tabular*}
\begin{tablenotes}
\item[*] The reference coordinates of V627 Cas used in the observations were R.A.=22:57:40.99 and Dec.=58:49:12.50 (J2000 obtained by IRAM Observation Logs (IRAM 1991-2015)).
\end{tablenotes}
\end{threeparttable}
\end{center}
\end{table*}

\subsection{Issue related with the distance to V627 Cas}
At present, the distance to V627 Cas is not well-established. \citet{coh1977} estimated the distance to be $\sim$3.3 kpc using the V-band extinction and absolute magnitude. However, \citet{ber1988} estimated it to be less than 800 pc. In addition, the parallax of V627 Cas was reported to be $\sim0.18\pm6.20$ mas in Hipparcos data \citep{van2007}, corresponding to a distance of about 5.56 kpc. While, GAIA data release 2 \citep{gaia2018}, showed the fit for parallax to be $\sim0.035\pm0.156$ mas, corresponding to a distance of 28 kpc. Using the distance converted from the parallaxes of the Hipparcos and GAIA catalogs, the physical scales for the ring radii of the SiO masers were estimated to be 20 AU and 100 AU, respectively. These values are too large compared with those of other Mira variables, for instance, which have a radius of 3.47 AU for R Aqr and 4.84 AU for R Cas \citep{min2014, as2011}. 

However, the errors of parallax in catalogs are large, and GAIA astrometry in DR2 has accuracy problems for binary systems \citep{xu2019}. Therefore, we presumed the distance of V627 Cas on the basis of the ring fitting results of SiO masers. If V627 Cas had the typical SiO maser ring size for red giants, $\sim 4$ AU, its distance would be $\sim 1.14$ kpc, which is inconsistent with any of the distance estimates appearing in the references mentioned above. Among them, the estimation by \citet{ber1988} to be $<$ 800 pc is the closest. This estimation will be compared with the parallax provided by GAIA catalog in coming releases, in which the parallax measurement for binary stars is expected to be improved.

\section{SUMMARY}

The results of simultaneous VLBI observations of $\rm H_2O$ $6_{1,6}-5_{2,3}$ (22 GHz), and SiO v=1, 2, J=1$\rightarrow$0, SiO v=1, J=2$\rightarrow$1 masers toward the suspected D-type symbiotic binary star V627 Cas made with KVN are presented, based on five epochs of data. Astrometrically registered maps of the $\rm H_2O$ and SiO masers in the epochs present a very asymmetric morphology and characteristics, differing from those of isolated single red giant stars. We performed the ring fitting for the SiO v=1, and v=2 maser features and determined the expected position of the red giant. The $\rm H_2O$ maser showed a very asymmetric feature with an one-sided conical outflow from epoch 3 with respect to the estimated position of the red giant. In addition, the $\rm H_2O$ maser features showed variations from north-northwest to west according to epochs 1, 3 and 5. Two possible scenarios of a bipolar outflow and the influence of the hot companion white dwarf in V627 Cas were discussed as the causes of the asymmetric distributions of $\rm H_2O$ and SiO maser lines. This is because the first scenario, a bipolar outflow, would disrupt the velocity coherence whereas in the second scenario, the influence of the hot companion white dwarf, UV photons from the white dwarf can dissociate the $\rm H_2O$ molecules in the face-on region of the red giant. 
Through the ring fitting, and assuming that V627 Cas has the typical SiO maser ring size of red giants, we derive the distance of V627 Cas to be about 1 kpc.

\section*{Acknowledgements}

This research was supported by a Major Project Research Fund (2018 - 2020) of KASI (Korea Astronomy and Space Science Institute) and by Basic Science Research Program through the National Research Foundation of Korea (NRF) funded by the Ministry of Education (2019R111A1A0105900). We are grateful to all of the staff members at KVN who helped to operate the array and the single dish telescope and to correlate the data. The KVN is a facility operated by KASI, which is under the protection of the National Research Council of Science and Technology (NST). The KVN operations are supported by KREONET (Korea Research Environment Open NETwork), which is managed and operated by KISTI (Korea Institute of Science and Technology Information).











\bsp	
\label{lastpage}
\end{document}